\documentclass[12pt,preprint]{emulateapj}
\usepackage{psfig}

\usepackage{graphicx} 

\slugcomment{Submitted to ApJ}

\shorttitle{Submm clustering}
\shortauthors{Blain et al.}

\begin{document}


\title{Clustering of Submillimeter-Selected Galaxies} 


\author{A.\,W. Blain and S.\,C. Chapman}
\affil{Caltech,
    Pasadena, CA 91125}

\author{Ian Smail}
\affil{Institute for Computational Cosmology, University of Durham, Durham, UK}

\and 

\author{Rob Ivison\altaffilmark{1}}
\affil{UK-ATC, Royal Observatory, Blackford Hill, Edinburgh, UK}
\altaffiltext{1}{Institute for Astronomy, University of Edinburgh, Blackford Hill,
Edinburgh, UK}



\begin{abstract}
Using accurate positions from very deep radio observations to guide
multi-object Keck spectroscopy, we have
determined a
substantially complete redshift distribution for very luminous,
distant submillimeter-selected galaxies (SMGs). A sample of 73
redshifts for SMGs in seven fields contains a surprisingly large number of
`associations': systems of SMGs with Mpc-scale separations, and 
redshifts within
1200\,km\,s$^{-1}$. This sample provides tentative evidence of
strong clustering of SMGs at $z \simeq 2-3$ with a
correlation length 
$\sim (6.9 \pm 2.1)h^{-1}$\,Mpc, using a simple pair-counting
approach that is appropriate to the small, sparse SMG samples. This is 
somewhat greater than the well-determined 
correlation lengths for both $z \simeq 3$ 
optical--UV color-selected
Lyman break galaxies (LBGs) and $z \simeq 2$ QSOs. This 
could indicate that SMGs
trace the densest large-scale structures in the
high-redshift universe, and that they may either be 
evolutionarily distinct from
LBGs and QSOs,
or subject to a more
complex astrophysical bias.
\end{abstract}


\keywords{galaxies: clusters: general -- galaxies: evolution -- 
galaxies: formation  
-- galaxies: starburst -- 
large-scale structure of universe -- 
cosmology: observations
}


\section{Introduction}

Since 1997 several hundred galaxies have been discovered using imaging 
arrays at millimeter and submillimeter wavelengths (Blain et al. 2002; 
Smail et al. 2002; Scott et al. 2002; Borys et al.\ 2003; Webb et al.\ 2003;  
Greve et al. 2004a). The difficulties of identifying these galaxies at 
other wavelengths, and moreover determining their redshifts, are 
significant, owing both 
to the coarse positional accuracy ($\simeq 8"$) of the discovery 
images and to the faint optical magnitudes of their proposed counterparts. 
By exploiting the high-redshift analog of the far-infrared--radio correlation 
between hot dust and radio synchrotron emission, which are 
powered by young, 
hot stars and their associated supernovae respectively, 
sub-arcsecond radio positions can be found for a substantial  
fraction of submillimeter-selected galaxies (SMGs). This 
enables efficient multi-object optical 
spectroscopy to search for 
redshifts, given that the surface 
density of radio-detected 
SMGs on the sky is $\sim 300$\,deg$^{-2}$ (Chapman et al. 2003a), 
and thus is well matched to the field of view and  multiplex gain of current 
spectrographs. This 
technique was first tried using the Keck LRIS-B spectrograph, which 
counterintuitively detected identifiable restframe ultraviolet spectral 
lines and features in many of these galaxies. 
The redshift distribution of a large sample of 73 of these galaxies in 
seven independent fields listed in Table\,1 is presented by 
Chapman et al.\ (2004a). The spectroscopic completeness for 
redshift determinations is
$\sim 50$\% for SMGs selected purely on the grounds of submillimeter 
flux density greater than 5\,mJy at a wavelength of 850\,$\mu$m, 
and $\sim 70$\% of for SMGs with the same submillimeter flux density and radio 
detections at 1.4\,GHz brighter than 30\,$\mu$Jy. 

The remaining $\sim 30$\% of radio-detected SMGs 
without redshifts are likely split   
between those at higher redshifts and any with cooler 
dust temperatures and lower luminosities, neither of which yield 
radio detections, and galaxies in the 
redshift range $1.2 < z < 1.8$, whose optical continua are generally too 
faint for the  
detection of rest-frame UV absorption lines, and for which  
any restframe UV emission lines are not redshifted into the wide spectral 
range of LRIS 
(Chapman et al. 2004a). 

This reasonably complete redshift distribution confirms absolutely that the 
SMGs are an important high-redshift galaxy population, and 
that galaxies within a factor of 5 in submillimeter flux density of the 5-mJy 
850-$\mu$m detection limit of this sample 
dominate the luminosity density from galaxies at redshifts $z \simeq 2$. 
The redshift distribution can be represented adequately by a Gaussian with 
$\bar z = 2.4$ 
and $\sigma = 0.65$, with almost all SMGs found over the redshift range 
$1.5 < z < 3$. SMGs are 
found in relatively small fields on the sky, 
less than 10' across, but are  
spread very widely in redshift from at least 0.8 to 3.4. 
This means that the SMG survey volumes have a uniquely long 
thin pencil geometry, which extends 
in depth by 2--3\,Gpc, but 
across the sky only as far as 5\,Mpc. 

The spatial distribution of SMGs 
is important, as the detection of any clustering signal can provide 
information about
the distribution of galaxies as a function of
dark matter density. In principle, in a dark
matter-dominated universe the strength of clustering should correlate with the 
mass of halos (Bardeen et al.\ 1986), and thus its measurement 
provides an insight into the masses
of the dark halos that host luminous galaxies 
(e.g. Overzier et al.\ 2003 and references therein).

The clustering strength of many types of high- and low-redshift
galaxies has been probed 
by a wide variety of methods.
At redshifts less than those of SMGs, $z<1.3$, the  
large-scale structure of galaxies is being mapped in great detail by 
the DEEP2 survey (Coil et al.\ 2003). At higher mean 
redshifts, the clustering of many hundreds of restframe 
UV/optically-selected, spectroscopically-confirmed 
Lyman-break galaxies (LBGs) 
can be measured (Adelberger et al. 1998). Recent values of correlation 
lengths\footnote{The
two-point 
correlation function of galaxies that describes the excess probability 
of finding a pair of galaxies separated by $r$ over an unclustered 
distribution is observed to be a decreasing power-law with distance
$\xi(r) = (r/r_0)^{\simeq-1.8}$ (Coil et al. 2003).} are
$r_0 = 4h^{-1}$\,Mpc (Porciani \& Giavalisco 2002) and $r_0 = 
2.7h^{-1}$\,Mpc (Ouchi et al. 2001). The published samples of 
LBGs lie at higher redshifts than 
most SMGs, 
which could on its own explain the lack of observed 
correlation between the positions of LBGs and SMGs (Peacock et al.\ 
2000; Chapman et al.\ 2000; Webb et al.\ 2003), 
even if they are drawn from a common underlying population of 
galaxies described by the same form of evolution. 
Lower-redshift counterparts to LBGs can be isolated 
by varying the color selection conditions (see
Steidel et al. 2004). Spectroscopically-confirmed samples of 
these galaxies can be compiled in parallel to surveys for 
SMG redshifts using the same multi-object slitmasks,    
and should ultimately provide an excellent 
opportunity to determine the relative three-dimensional distribution of  
LBGs and SMGs. 
The clustering of much rarer QSOs and relatively faint 
($\simeq 10$\,mJy) radio galaxies at redshifts that overlap with the 
SMGs has been probed by Croom et al. (2002) and Overzier et al. (2003), 
who report correlation lengths 
$r_0 = 5$ and 6$\,h^{-1}$\,Mpc respectively.
However, these surveys are more sparsely sampled, and cover larger areas of 
sky (of order thousands of square degrees), so the
typical source separation is much greater than the correlation  
length.

Without redshifts to trace three-dimensional structure, only limits to
the projected clustering strength of SMGs
in two dimensions have been obtained (Carilli et al. 2000; Scott et al.
2002; Borys et al. 2003; Webb et al. 2003; see Fig.\,1). This is unsurprising
given the narrow, deep geometry of SMG surveys:
multiple structures along the line of sight overlap
and dilute the clustering signal.
Note that even the best possible photometric redshifts derived from 
deep multicolor optical and near-IR imaging with $\Delta z \simeq 0.1$ 
are not sufficiently accurate to provide useful three-dimensional clustering 
information in the 
deep, narrow pencil-beam geometries of SMG surveys, which include very 
significant projection effects (see Fig.\,2). 
The availability of a large number of spectroscopic redshifts 
means that we can now remove these projection effects and investigate the 
clustering of the SMG population directly. 

Throughout this paper we assume a post-{\it Wilkinson Microwave Anisotropy 
Probe}
cosmology with $\Omega_0=0.27$, $\Omega_\Lambda=0.73$, and 
$H_0=100h$\,km\,s$^{-1}$\,Mpc$^{-1}$ with $h=0.71$.  

\vspace{12cm} 
\begin{center}
\includegraphics[width=.3\textwidth,angle=-90]{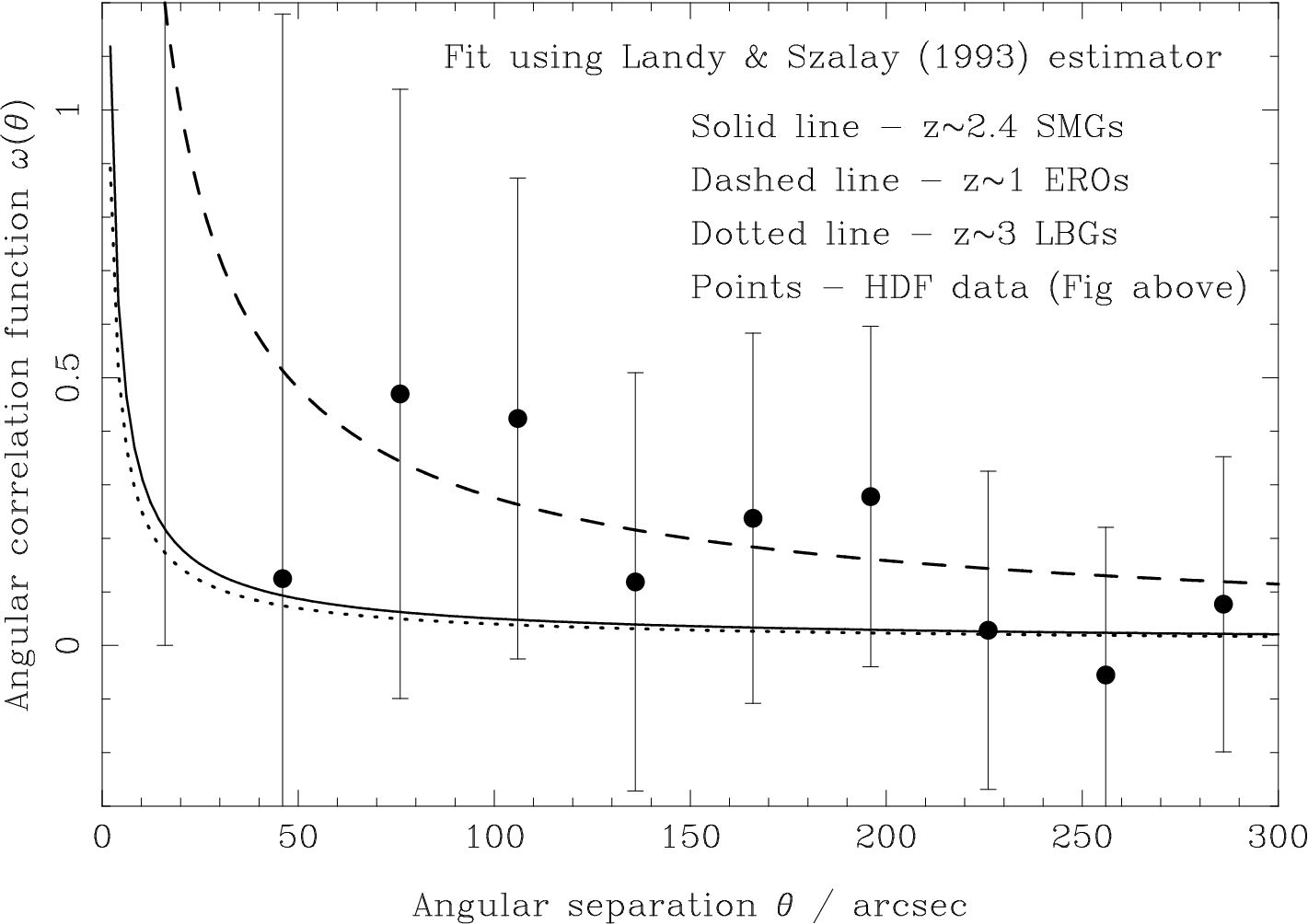} 
\end{center} 
{\footnotesize{{\sc Fig. 1. ---} 
ACF observed for  
47 SMGs in the extended HDF field shown in Fig.\,1 (Borys et al.\ 
2003; Chapman et al.\ 2003a, 2004a). Other results 
for two-dimensional clustering of SMGs have been discussed by 
Carilli 
et al. (2000), Scott et al.
(2002), Borys et al. (2003) and Webb et al. (2003). 
The errors on existing ACF data are much too great 
to measure the clustering signal without much larger SMG samples: 
redshifts 
are essential to discern the clustering of SMGs.
The observed correlation functions 
of $z \simeq 1$ EROs (Daddi et al. 2001; $r_0 \simeq 12h^{-1}$\,Mpc), 
$z \simeq 3$ LBGs (Adelberger et al.\ 1998; Porciani 
\& Giavalisco 2002; $r_0 \simeq 4h^{-1}$\,Mpc) 
and $z \simeq 2.5$ 
SMGs assuming the correlation length $r_0 \simeq 7h^{-1}$\,Mpc  
determined in Section\,3 (Table\,1) 
are also shown.   
After projection, the SMGs and LBGs have similar ACFs, which are  
much weaker than the ACF for the lower-redshift 
EROs.}}
\label{fig1}

\section{Signs of clustering in SMG surveys} 

There have been serendipitous detections of SMGs clustered 
in close proximity to 
other classes of high-redshift galaxies. For example there are 
detections of optically-selected galaxies at redshifts matching those 
of known SMGs and with separations  
of only $\sim 100$\,kpc in 
the backgrounds of two gravitational lensing clusters 
of galaxies:
at $z=2.80$ in A370 (Ivison et al.\ 1998; 
Santos et al.\ 2004), and at $z=2.55$ in 
A2218 
(Kneib et al.\ 2004). SMGs were identified by Chapman et al. (2001) 
in the most overdense structure of LBGs at $z \simeq 3.1$ (Steidel 
et al.\ 2000). 
Overdensities of bright SMGs found near the targeted 
positions of 
some of the most extreme high-redshift radio galaxies at redshifts 
as great as $z=3.8$ (Ivison et al. 2000; Stevens et al. 2003) also 
provide evidence for some strong overdensities in the 
spatial distribution of SMGs, some confirmed by spectroscopy 
(Smail et al.\ 2003a, b). 
Can we find direct spectroscopic evidence of clustering in the general 
population 
of SMGs? 

In order to demonstrate the crucial importance of obtaining  
redshifts for the SMGs before attempting to measure their clustering 
properties,  
in Fig.\,1 we determine the projected two-dimensional angular 
correlation function (ACF)  
of 47 SMGs in the 
biggest available survey field (Borys et al.\ 2003; Chapman et al.\ 2004a).
To measure a significant ACF for the SMGs the large 
fractional errors on the results shown in Fig.\,1 
indicate that a {\emph{much}} larger sample of several hundred SMGs would be 
required, a challenge for the next generation of sensitive wide-field 
submillimeter telescopes. 

We indeed find spectroscopic evidence for clustering of SMGs. 
In six of the seven fields containing SMGs with redshifts,
"associations" of galaxies are found, with redshifts separated by less than 
1200\,km\,s$^{-1}$ in radial velocity (Table\,1). This velocity is an 
approximate upper limit to the velocity dispersion of the richest clusters of 
galaxies at the present epoch, and so should be larger than the velocity 
dispersion of any less evolved large-scale structures at high redshifts. 
Furthermore, it is greater than the random 
offset velocity expected from 
emission generated in randomly oriented outflowing 
galactic winds at high-redshifts 
(Erb et al. 2003). 
Most of the associations are pairs, but 
there is a group of three SMGs 
at $z=3.1$ in the highest overdensity in the survey 
of LBGs in the SA22 field (Steidel et al. 2000), 
and a group of five SMGs at $z=2$ in the Hubble Deep Field (HDF; Fig.\,2). 
SMGs are often found to have 
disturbed and probably interacting morphologies (e.g. Ivison et al. 1998; 
Chapman et al. 2003b). Note that the associations 
are not close interacting pairs of SMGs (such as, e.g., 
SMM\,J09431+4700; Neri et al.\ 2003; see also Swinbank et al.\ 2004), 
but are 
separated on much larger scales up to several Mpc. Their spatial separation 
is comparable both to the extent of the 
survey fields (Fig.\,2) and to the scale of the local 
galaxy correlation length. It is possible  
that the successors of the SMGs may finally reside in 
the densest clusters, but at the observed epoch the SMGs with 
associated redshifts are much more 
widely spread, presumably lying throughout infalling, pre-virialized 
filaments and walls in the large-scale structures of galaxies. 

In parallel to measuring the SMG redshifts, Chapman et al.\ (2004b) have found 
redshifts for a comparable number of optically-faint radio galaxies (OFRGs) 
that are not detected at submillimeter wavelengths. Their optical spectra 
are similar to the SMGs, and it is very plausible that these submillimeter-faint 
OFRGs are 
comparably luminous to the SMGs, but have dust temperatures too high to 
be detected at submillimeter wavelengths (Blain et al. 2004). 
The submillimeter-faint OFRGs display a similar 
abundance of systems with associated redshifts to the SMGs, both 
within the submillimeter-faint OFRG  
population itself (in 
the 03 hr, Lockman, SA13 and SA22 fields), and
when cross-correlated with the SMG population 
(in the five-member HDF association 
and in the very overdense $z=3.1$ structure in the SA22 field). 
Once {\it Spitzer Space Telescope} observations 
reveal whether these galaxies are truly as luminous as the SMGs, they 
offer to approximately double the density of the 
sampling of large-scale structure 
provided by very luminous dusty galaxies. 

\setcounter{figure}{1}
\begin{figure*}[t]
\begin{center}
\includegraphics[width=.37\textwidth,angle=-90]{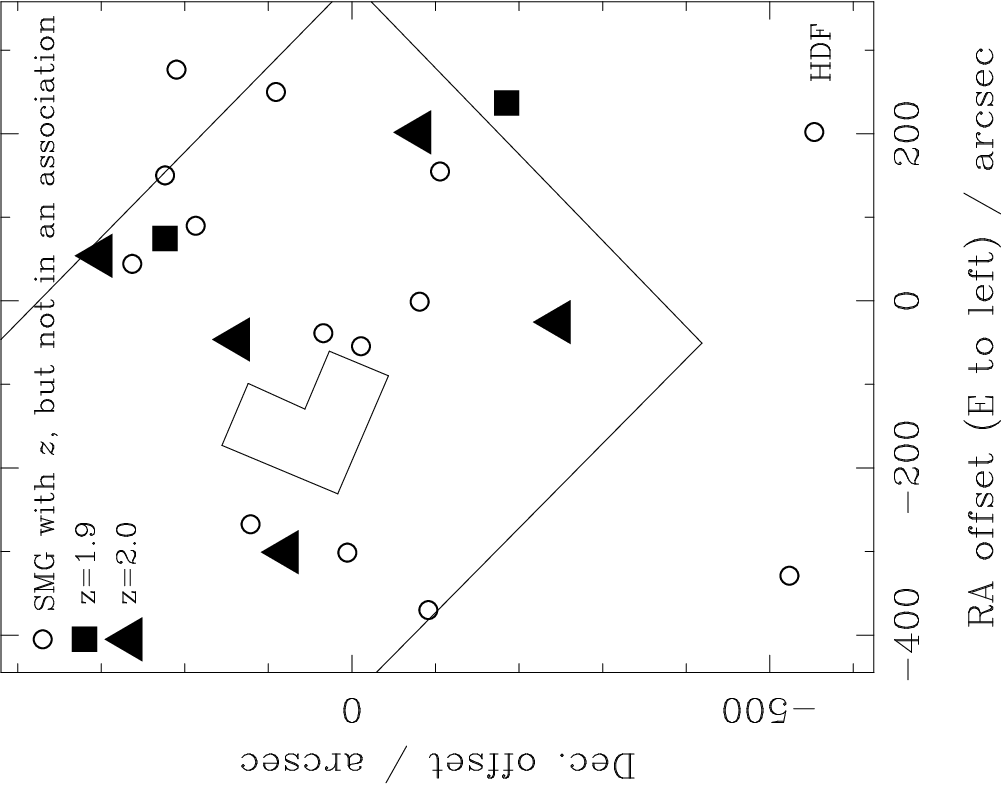}
\hspace{10pt}
\includegraphics[width=.37\textwidth,angle=-90]{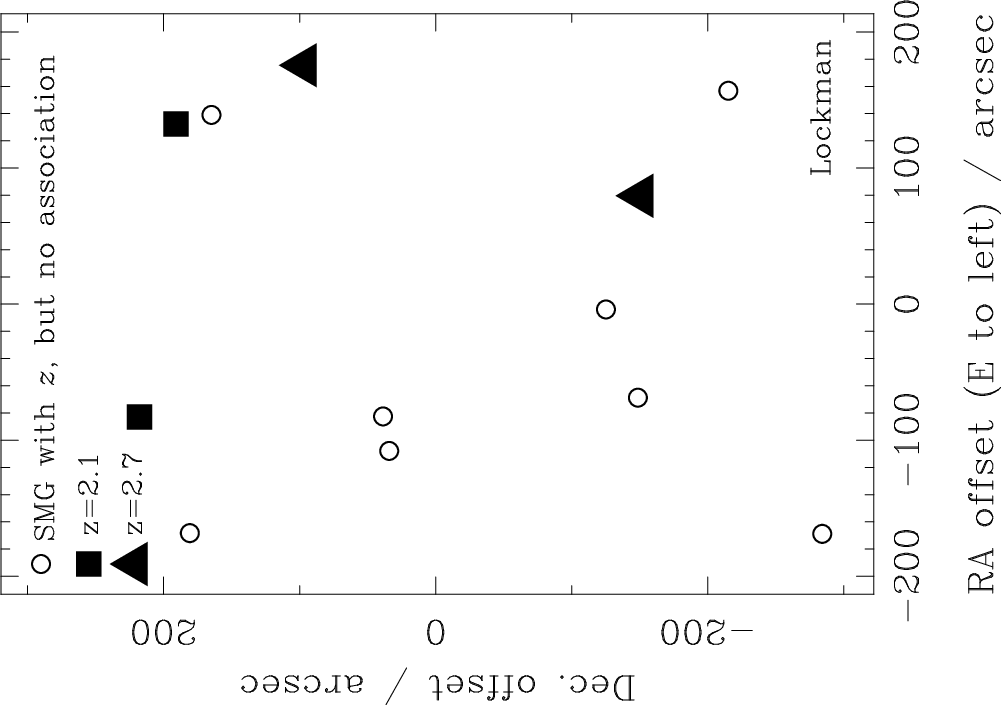} 
\hspace{10pt} 
\includegraphics[width=.37\textwidth,angle=-90]{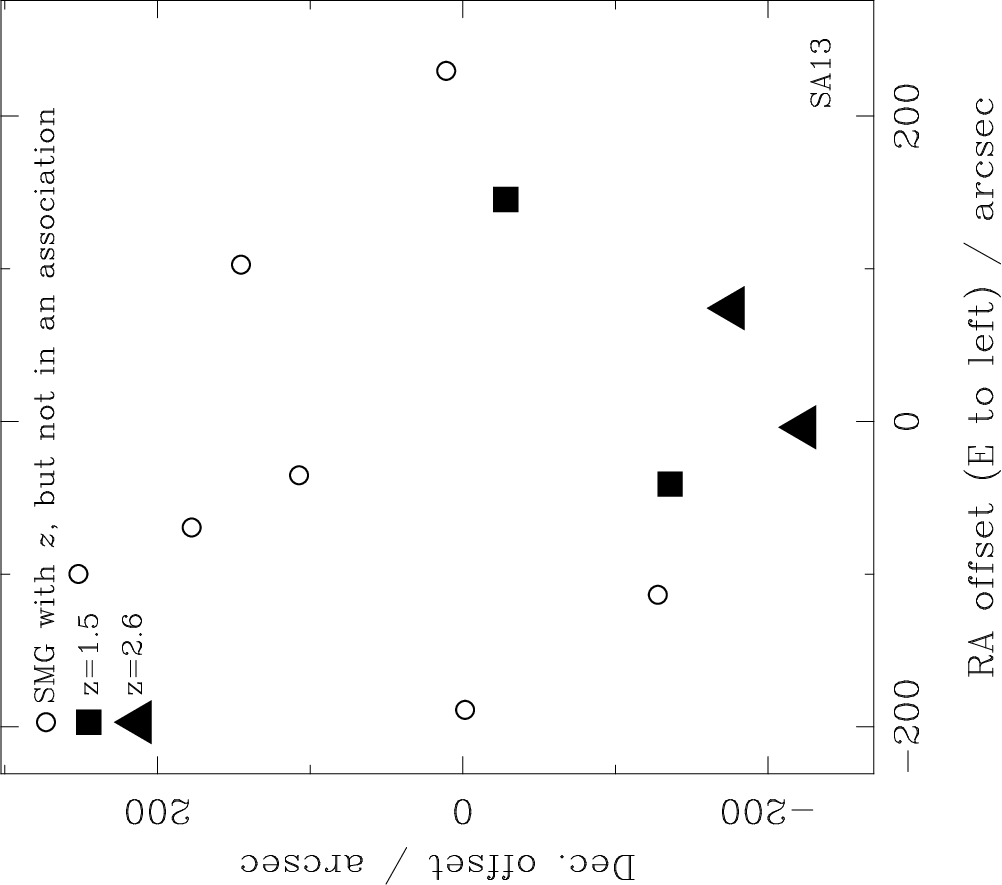} 
\end{center}
\caption{
Positions of radio-pinpointed SMGs in the 
richest three field in our redshift survey: from left to right these are;
the HDF (the boundaries 
of the HDF-N and GOODS fields are also shown), Lockman Hole, and SA13 fields. 
SMGs in our catalog 
with spectroscopic redshifts are represented by open circles. Those 
in the redshift `associations' 
present are shown by different filled symbols. These
figures demonstrate the typical wide separation of the 
associations, which span the whole field. 
}
\label{fig2} 
\end{figure*} 

\section{The correlation length of SMG clustering}  

To interpret our sparse data, we cannot use regular methods to 
determine clustering strength (Adelberger et al.\ 1998; Wilson et al.\ 
2001) but 
must make use of the limited information as best we can. 
We compare the 
number of pairs of galaxies found in each field with velocity 
separations less than 1200\,km\,s$^{-1}$ (Table\,1) with 
the number of pairs 
expected assuming a comoving correlation length that is fixed 
throughout the 
redshift range of the SMG sample. We further assume that the  
redshift distribution of the population is known accurately, 
and determine the 
space density of SMGs in each field that is required 
to match the observed number. This allows us to estimate the number 
of pairs expected in each field in the absence of clustering, as 
listed in Table\,1 ($N_{\rm pair}'$).  
To evaluate the number of pairs expected with a particular  
correlation function $\xi$, which is assumed to be constant in redshift,  
we carry out a double integral over volume  
for the product of the space density of SMGs $N$ at two positions 
separated by a three-dimensional distance $r$, and the 
two-point correlation function $\xi$:
\begin{eqnarray} 
N_{\rm pair} &=& {1\over2} \int_{\theta, \phi} \int_{\theta', \phi'} 
\int_{z_{\rm min}}^{z_{\rm max}}  
\int_{-\Delta v}^{\Delta v} N(z,\theta,\phi) \times \\ 
& & \>\>\>\>\>\>\>\>\>\>\>\>\>\>\>\>\>\> N(z',\theta',\phi') \times  
\left[ 1 + \xi(r) \right] {\rm d}V {\rm d}V'.
\end{eqnarray} 
The factor of a half corrects for double counting. 
One radial variable is limited by the redshift range of the SMGs 
(values of $z_{\rm min} = 1.8$ and $z_{\rm max}=4.0$ were used, but 
note that the effects of changing $z_{\rm min}$ to zero are less than 1\%:
the peaked distribution of $N$ sets the size of $N_{\rm pair}$), 
and the other the distance corresponding to 
relative velocity in the Hubble flow, with a limit $\Delta v$ 
set to the limiting velocity that defines a pair. This is usually   
1200\,km\,s$^{-1}$.  
Note that we integrate over the velocity 
range as we cannot be sure of the relative spatial distribution of 
galaxies along the line of sight because of peculiar velocities, and so 
all galaxies that lie within $\Delta v$ of each other are counted. 
The solid angle of the survey in each rectangular field sets the 
limits on $\theta$ and $\phi$.
$N$ is assumed the 
have the Gaussian redshift distribution discussed above, normalized to 
match the number of detections $N_{\rm gal}$ in each field (Table\,1). 

We then change the correlation length $r_0$ included in $\xi(r)$,  
keeping $\gamma = -1.8$, until 
the numbers of pairs predicted by equation 1 matches the 
observed numbers of pairs $N_{\rm pair}$ (Table\,1), either 
field by field or in a combination 
of fields, as listed in Table\,1. The uncertainties on the 
derived values 
of $r_0$ are determined by the range of $r_0$ values that match the 
Poisson errors on $N_{\rm gal}$ and $N_{\rm pair}$ in each field. 
$N_{\rm pair}'$ in  Table\,1 is derived for $r_0 = 0$\,Mpc.  
Note that the overall result is presented with and without the 
SA22 field included, as the overdensity in that field was known 
prior to substantial submillimeter observations, although the radio-pinpointed 
submillimeter observations 
were carried out before any redshifts were determined. 

We now illustrate the robustness of this pair-counting procedure in an 
LBG survey field with a more richly sampled redshift 
distribution.
The largest and best populated LBG survey field -- the 
850"$\times$850" Westphal  
field\footnote{ 
ftp://ftp.astro.caltech.edu/pub/ccs/lbgsurvey} -- 
contains 
192 galaxies with redshifts, and 912 galaxy pairs using our definition.  
As the LBG redshift distribution is approximately a Gaussian 
with $\bar z = 3.0$ and $\sigma_z = 0.28$, 594 pairs would be expected with 
no clustering. This corresponds to an excess of pairs over random 
by a factor of 
1.55, which requires a correlation length of 
$r_0=(5.8 \pm 0.3)h^{-1}$\,Mpc using our 
method of integrating over the two-point correlation function. 
By comparison, we obtain a correlation length $r_0=(6 \pm 0.3)h^{-1}$\,Mpc 
in this field if we
evaluate the two-point ACF 
against a mock catalog 
of random galaxy positions with the same redshift distribution 
(Landy \& Szalay 1993), and then compare the results with the 
ACF obtained from an assumed three-dimensional 
correlation function using Limber's equation 
e.g. Wilson 2003). Although this is a single relatively small 
field, and fluctuations 
in the correlation amplitude are certainly expected, the result 
from the pair-counting method used to analyze our small SMG sample 
is in accordance 
with that expected from more established methods.  Note that the 
strength of clustering in this field appears to be greater than 
for the LBG sample as a whole. 

Over all seven fields we find a correlation length 
$r_0 = (6.9 \pm 2.1)h^{-1}$\,Mpc for the SMG sample. 
Only in the best-sampled HDF field is a significant result
obtained in an individual field -- $r_0 = (9.5 \pm 3.3)\,h^{-1}$\,Mpc 
-- the other
fields all include too few galaxies and pairs.  
Note that varying the choice of field area or $\Delta v$ 
makes relatively little difference to the result. 
If the size of the HDF 
field is doubled, then the calculated correlation length is 
reduced to 
$r_0 = (7.5 \pm 3.2)h^{-1}$\,Mpc, consistent with the value above. 
If $\Delta v$ is reduced to 600\,km\,s$^{-1}$, then only two pairs 
of galaxies are lost from the 18 in the sample. The calculated 
correlation length is then increased by 24\%. If the 
value of $\Delta v$ is increased to 1800\,km\,s$^{-1}$, then 
an additional 
three pairs are counted, and the correlation length is reduced by 
8\%. Hence, within the uncertainties the results are robust to changes in 
the assumptions made.   

The correlation length for SA22 is greater, but note that SA22 includes
a known strong overdensity at the redshift of the association of SMGs
(Steidel et al. 2000), and so it is possible that  
the value for 
SA22 does not give a fair representation of the clustering of SMGs in a 
typical field. 

Using the larger spectroscopic catalogs of LBGs, 
we also conducted several additional tests of the results. We again
verified that the predicted and measured counts of pairs are 
not very sensitive to reasonable changes in the chosen  
velocity range $\Delta v$ away from 1200\,km\,s$^{-1}$. The results 
depend more strongly on 
the shape and size of the field, with smaller fields yielding a 
larger boost in the expected number of pairs for a certain correlation  
length, as the correlation function $\xi$ is larger when 
sampled on smaller scales. We also 
randomized the redshifts of the LBG catalog and, as expected, 
found that the measured excess of pairs vanished. The 
pair-counting method is thus sensitive to real three-dimensional 
structure in the galaxy 
distribution. We also examined  
the effects of randomly dropping a redshift-independent 
fraction of the LBGs from the 
catalog, to simulate 
the effects of incompleteness in sampling a galaxy population 
with an well-known and independent correlation length 
measurement. The 
results do not change significantly, even if the sample becomes 50\% 
incomplete. As we know that the catalogs of redshifts for 
radio-selected SMGs are 
probably complete at the 70\% level (Chapman et al.\ 2004a), 
and that no redshifts in the range 
$1.8 < z < 4$ are preferentially excluded from our redshift sample, 
we can be confident in the 
accuracy of our results. 

A correlation length of $\sim 7\,h^{-1}$\,Mpc for the SMGs is close to 
that predicted by 
Shu, Mao \& Mo (2001), assuming that star-formation activity in 
massive halos follows a Schmidt Law with gas surface density; 
this is a very 
uncertain assumption given the apparent large masses, and complex 
dynamical structure in these galaxies (Frayer et al.\ 1998, 1999; 
Neri et al.\ 2003; T. R. Greve et al.\ 
2004b, in preparation). 
We can also compare and 
contrast the correlation length result for the SMGs with the 
results for rarer, 
more sparsely sampled populations of galaxies. The sample of QSOs with a 
surface density of $\sim 30$\,deg$^{-2}$ and 
$z \simeq 1.6 \pm 0.65$ from the Two-Degree Field (2dF) survey 
overlaps in redshift with the lower end of the SMG redshift 
distribution. The correlation function of the 2dF QSOs is smaller than 
that for the SMGs, 
$\sim 5\,h^{-1}$ Mpc (Croom et al.\ 2002). The 2dF correlation function is 
evaluated over a very much larger 
area ($\sim 730$\,deg$^{2}$), and so is not 
prone to field-to-field variations. The correlation length for 
other samples of high-redshift 
galaxies, including very 
deep near-IR selected populations (Daddi et al.\ 2003)  
and bright radio-galaxies (R\"ottgering et al. 2003) have also been evaluated 
at similar redshifts to our SMG samples. For the former, the survey areas are 
small in extent, and have relatively 
incomplete redshift distributions, while for the latter, the survey volumes 
are sparsely sampled. It is hence
difficult to compare the correlation functions
of these samples directly with SMGs at present.  

We are targeting a statistically reliable  
subset of our SMGs with known redshifts to detect 
CO emission lines using 
mm-wave interferometry and rest-frame optical nebular line 
emission using near-infrared spectrographs. 
The CO linewidth can be used to 
provide a dynamically-determined estimate of the mass within about 10\,kpc 
(Frayer et al.\ 1998; Neri et al.\ 2003; Genzel et al.\ 2003; 
T. R. Greve et al.\ 2004b, in preparation); a more accurate 
mass can be obtained if the emission can be resolved spatially. Initial 
indications are that SMGs have velocity dispersions that are greater by a
factor of 2, and thus have dynamical masses greater by a factor of $\sim 4$,
as compared with the dynamical masses of 
LBGs at $z \simeq 2$ ($4 \times 10^{10}\,M_\odot$; 
Erb et al.\ 
2003). The half-light radii of SMGs appear to be larger than those of 
SMGs (Chapman et al.\, 2003b), which would lead to a further increase in 
the relative mass of SMGs as compared with LBGs by a factor of 2. 
At $z \simeq 2$ near-IR observations of resolved nebular line 
emission are easier, although published samples of $z \simeq 3$  
LBGs are larger. 
LBGs appear to be typically twice as 
massive at $z=2$ as at $z \simeq 3$. None of 
these direct dynamical measures indicate, however, that SMGs are 
as massive as the 
several $10^{13}\,M_\odot$ halos their clustering 
properties would suggest (Fig.\,2). Their  
central dynamical masses derived on 10\,kpc length scales 
are likely to be a 
factor of only 3--4 times less than the mass of the dark-matter halo, 
extended to of order 200\,kpc, given the steep $r^{-3}$ decline in the 
expected profile of dark matter halos outside of the visible 
galaxy (Navarro, Frenk \& White 1997). 
Recent near-IR and CO observations of 
SMGs (Genzel et al.\ 2003; Tecza et al.\ 2004; T. R. Greve et al.\ 
2004b, in preparation; Swinbank et al.\ 2004) tend to support the
greater masses of SMGs as compared with LBGs.  
It is interesting to note that the masses 
of the dark-matter halos surrounding 
the large, well-studied sample of 
LBGs at $z \simeq 3$ inferred from their clustering properties 
($\sim 8 \times 10^{11}\,M_\odot$; Adelberger et al.\ 1998; Fig.\,1)  
also exceed the central dynamical masses 
inferred from spatially 
structured asymmetric line profiles and velocity dispersions, but by 
a smaller factor than for the SMGs. If in fact SMGs can only form in the most 
massive halos, then in a hierarchical picture of galaxy formation, this 
could account for their apparent typical moderate redshifts 
(Chapman et al.\ 2003a, 2004).  

Differences between 
the inferred dynamical and dark matter masses of high-redshift galaxies 
provides a direct hint that more 
complex astrophysics in SMGs could break a simple 
link between clustering amplitude and 
mass (see Shu et al.\ 2001). For example, the time-dependence of 
merging has been claimed to 
both boost (from infall; Scannapieco \& Thacker 2003), and leave unaffected 
(Percival et al. 2003) the 
clustering properties of galaxies. Intuitively,  
very luminous SMGs that are only briefly visible 
in overdense regions might be expected to have a weaker
clustering signal, as compared with a quiescent population that has 
the same spatial distribution; 
however, the strength of clustering could 
be increased if the luminous SMG phase is 
triggered preferentially in overdense 
regions. Further observations to investigate the 
relationship between the luminous SMGs 
and their underlying dark matter distribution are necessary to  
account for the effects of time evolution. Note that there is 
also certain to be strong field-to-field variation in the density 
of structures of associated redshifts for SMGs (Fig.\,2).  

We stress that accurate spectroscopic redshifts are essential 
in order to reveal clustering properties of the SMGs. The same 
information cannot be derived from two-dimensional data. 
The expected ACFs for $z \simeq 1$ extremely red objects (EROs), 
$z \sim 3$ LBG, and SMGs assuming the correlation lengths discussed above 
are shown in Fig.\,1. 
The relatively narrow redshift range for the 
EROs leads to a large expected signal, while the LBGs and SMGs  
have a much weaker ACF due in part to the greater distance to 
both samples, and mainly to their  
broader redshift distributions, especially for the SMGs. 
The inferred correlation length for the SMGs is similar to the projected 
extent of the SMG survey fields, 
and so the relative 
spatial positions of the members of each association reveal 
little information on the maximum extent and strength of the 
clustering present. We await much larger square degree-scale fields in 
order to sample the clustering properties of SMGs in a representative 
region of the high-redshift universe.   

\begin{center}
\includegraphics[width=.55\textwidth,angle=-90]{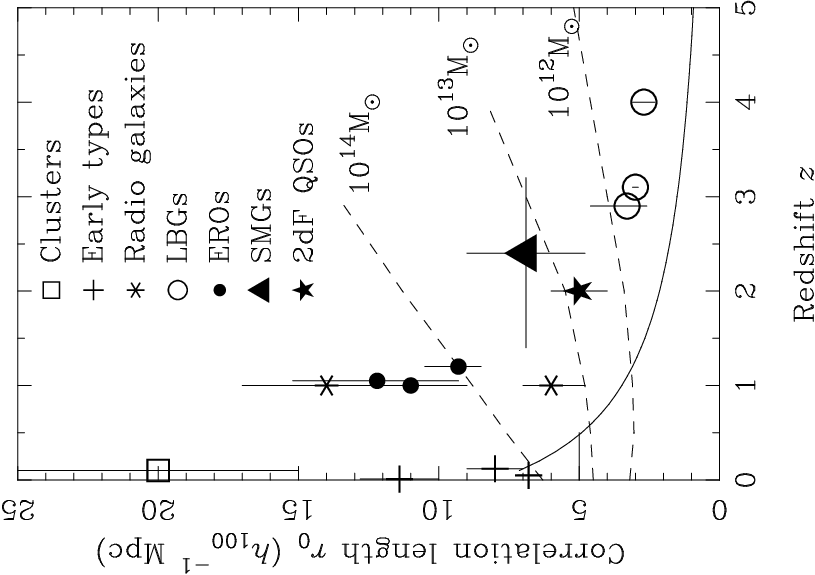} 
\end{center} 
{\footnotesize{{\sc Fig. 3. --- }
Comoving correlation 
length of the SMGs inferred from our survey in 
contrast to other populations of low and high-redshift galaxies (see
the summary in Overzier et al.\ 2003). The horizontal error bar on the 
SMG point spans the 
range of redshifts over which SMG associations are found (Table\,1). 
The solid line shows 
a representative model for 
the evolution of a certain overdensity. The dashed 
lines show the expected correlation length of dark 
matter halos as a function of mass and redshift. }}
\label{fig3}

\section{Consequences of strong SMG clustering} 

There are two immediate consequences of the presence of a strong clustering 
signal in the SMG survey. These concern both the nature of the SMG themselves, 
and the galaxies that result when the SMG activity is completed. 
First, since the correlation length of SMGs 
appears to be somewhat larger than for 
both LBG and QSO surveys at comparable redshifts, it seems  
unlikely that the SMGs form a simple evolutionary sequence with 
either population. It is 
attractive to link together these different samples of high-redshift galaxies,  
but they are likely to be substantially 
distinct populations. If they are distinct then the large luminosities of 
the SMGs would not represent a shorter-lived, more luminous 
phase during the active lifetime 
of more typical, less deeply dust-enshrouded star forming 
galaxies, and few powerful active galactic nuclei (AGNs) would 
burn out of the dust cocoons of SMGs to appear as optically visible QSOs. 
Second, if there are no complex biases at work, then the correlation 
length inferred for the 
SMGs is consistent with a form of evolution that 
subsequently matches the large comoving 
correlation length typical of evolved EROs 
at $z \simeq 1$ and of   
clusters of galaxies at the present epoch (Fig.\,3). 
If this picture 
is correct, then the 
descendants  
of SMGs would be 
rare in the field, and found predominantly in rich cluster environments. 
However, this appears to be inconsistent with their space densities, since a
progenitor of a rich cluster of galaxies is not 
expected to 
intersect every 100-arcmin$^2$ field sampled out to redshift 
$z \simeq 3$. The condition for this to occur is 
a comoving density of proto-clusters on the order of $10^{-6}$\,Mpc$^{-3}$, 
in contrast to the observed 
comoving density of clusters with masses 
$M > 8 \times 10^{14}\,M_\odot$ at $z=0$, which is 
$\sim 10^{-7}$\,Mpc$^{-3}$ (Bahcall et al.\ 2003). However, it is 
possible that SMGs in more extended unvirialized wall and filament 
large-scale structures with masses of several times 10$^{13}\,M_\odot$ 
at $z \simeq 2.5$ 
(Fig.\,3) that subsequently drain, merge and collapse into more 
massive clusters   
could provide a greater effective covering factor, to reconcile these
values. 
More likely, either the halo masses are less extreme than inferred from a
picture of simple one-to-one halo biasing at a given mass,
or some other astrophysical biasing effect must be at work. By considering the 
correlations between dark matter sub-halos, and a 
"halo occupancy distribution", 
it is possible to enhance clustering on a 
given halo mass scale 
(Berlind et al.\ 2003). Building a better observational view of 
the density of the environments of SMGs based on deep multiwaveband imaging 
and densely-sampled 
spectroscopy should allow these ideas 
to be tested.

The suggestion of an evolutionary link between
SMGs and passive EROs from their clustering strength 
may provide useful constraints on the lifetime
of the SMG phase.
The comoving space density of bright radio-detected 
SMGs at $z \simeq 2.5$ 
is $\sim 2.7 \times 10^{-5}$\,Mpc$^{-3}$. The 
corresponding density of EROs at 
$0.7<z<1.5$ is $(2.2 \pm 0.6) \times 10^{-4}$\,Mpc$^{-3}$ in a sample
with spectroscopic completeness of $\sim 70$\% (Cimatti et al.\ 2002). 
The density of EROs matches closely  
the comoving abundance of $L^*$ elliptical galaxies at the present epoch, 
which is 
$\sim 2 \times 10^{-4}$\,Mpc$^{-3}$. 
Hence, 
because the comoving abundance of SMGs is much less, even neglecting merging,   
the SMGs and EROs 
can only be part of a direct evolutionary 
sequence with the same density in space 
if the duty cycle 
of SMGs is short, less than 700\,Myr, to generate   
enough evolved, long-lived EROs in the 6\,Gyr prior to 
$z \simeq 1$. The resulting EROs could 
then evolve naturally into $L^*$ 
elliptical galaxies at the present epoch with a very modest amount of 
merging. The unknown time profile of the luminosity of SMGs would 
certainly also have an effect on the results.
Direct accurate 
determinations of the masses of SMGs
are critically important to investigate these links. 
The first hints from systematic 
CO observations are that SMGs 
are massive enough to generate the stellar populations of  
$L^*$ elliptical galaxies directly 
(Neri et al.\ 2003; T. R. Greve et al.\ 2004b, 
in preparation). 

The correlation function of SMGs appears to be consistent with their 
being associated with massive dark-matter halos at high redshifts, halos more 
massive than those hosting LBGs and QSOs at comparable redshifts. 
The successors of SMGs could thus be associated with clusters of galaxies at 
the present epoch. A key test of the nature of the SMGs requires 
a larger sample of 
dynamical mass estimates for the population, and a direct comparison 
of their spatial 
distribution as compared with populations of more common 
optical/UV-selected 
galaxies in the same fields. 

\section{Summary}

The discovery of associated systems of high-redshift 
luminous SMGs, enabled by the measurement of 
a substantially complete redshift distribution 
($z \simeq 2$--3),   
hint that they represent a strongly clustered population. We derive 
a correlation length of $(6.9 \pm 2.1) h^{-1}$\,Mpc in our best sampled 
field. Representative samples of 
high-redshift SMGs 
may thus be found in high-density environments that could evolve into rich 
clusters of galaxies at the present epoch. The large correlation length 
casts doubt on a simple evolutionary link with populations of 
high-redshift optical/UV-selected galaxies and QSOs, while offering the 
potential to use associations of SMG redshifts to signpost the densest 
regions of 
the high-redshift Universe with only a modest requirement on spectroscopic 
observing time. 
A key goal is now to investigate and compare the masses of SMGs inferred 
from clustering measurements, and the dynamical masses inferred from 
near-IR and mm-wave CO spectroscopy. This could help to determine the 
nature of the power source and lifetime of the SMGs themselves, and reveal the
galaxies at the present epoch that evolve from the SMGs. 

\acknowledgments



A. W. B. acknowledges support from NSF grant AST-0205937 and 
the Alfred P. Sloan foundation. I. R. S. acknowledges support from 
the Royal Society. We thank the anonymous referee for 
their helpful comments on the manuscript, and to Kurt Adelberger for 
pointing out a typo in equation 1.







\begin{deluxetable}{lllclllll}
\tablecaption{
Correlation Lengths $r_0$ for Radio-pinpointed SMGs.
\label{tbl-1}
}
\tablewidth{0pt}
\tablehead{
\colhead{Field} & \colhead{R.A.} & \colhead{Decl.}  &
\colhead{Size} & \colhead{$N_{\rm gal}$} & 
\colhead{$N_{\rm pair}$} & \colhead{$z_{\rm as}$} & \colhead{$N'_{\rm pair}$}
& \colhead{$r_0$} \\
 & & & \colhead{(arcsec)} & & & & & \colhead{($h^{-1}$\,Mpc)}
} 

\startdata
All fields& ... & ... & ... & ... & ... & ... & ... & $6.9 \pm 2.1$ \\
All but SA22& ... & ... & ... & ... & ... & ... & ... & $5.5 \pm 1.8$ \\
03hr & 03 02 & 00 06 & $240 \times 280$ & 5 & 0 & ... & 0.20 & ... \\
Lockman & 10 52 & 57 20 & $360 \times 450$ & 12 & 2 & 2.1, 2.7 & 1.2 & $4.3 \pm 3$\\
HDF & 12 36 & 66 10 & $720 \times 900$ & 21 & 9 & 1.9, 2.0\tablenotemark{a} & 3.65 & 
$9.5\pm3.3$ \\
SA13 & 13 12 & 42 40 & $470 \times 540$ & 11 & 2 & 1.5, 2.6 & 0.99 & $6.1 \pm 4$\\
14hr & 14 18 & 52 29 & $320 \times 210$ & 7 & 1 & 2.1 & 0.40 & $5.3 \pm 5.3$ \\
N2 & 16 37 & 40 55 & $480 \times 750$ & 8 & 1 & 2.4 & 0.53 & $6.4 \pm 6.4$\\
SA22 & 22 18 & 00 12 & $820 \times 650$ & 9 & 3 & 3.1$\tablenotemark{b}$ &
1.0 & $15 \pm 9$ \\
\enddata
\tablecomments{
The total number
of spectroscopic redshifts $N_{\rm gal}$ and the numbers of
pairs of galaxies
$N_{\rm pair}$ within
1200\,km\,s$^{-1}$ in radial velocity at redshift $z_{\rm as}$ in each field 
are listed, along with
the pair abundance predicted
in the absence of clustering $N'_{\rm pair}$. Units of right 
ascension are hours and minutes, and units of declination are 
degrees and arcminutes. 
}

\bigskip 

\tablenotetext{a}{This is a 5-galaxy association} 
\tablenotetext{b}{
These three SMGs lie in the richest
spectroscopically confirmed high-redshift large-scale
structure (Steidel et al.\ 2000), which contains 29 of the 76 LBGs
in this field. If the width of the LBG redshift distribution 
matched that of the SMG surveys, then only 15\% of
the LBGs
would lie in this structure, not the
3/9 found in the SMG survey. The comoving density of LBGs in the 
best-sampled Westphal--14\,hr field is $4.6 \times 10^{-4}$\,Mpc$^{-3}$, 
while the comoving density of SMGs in all the survey fields is 
$(2.7 \pm 0.4) \times 10^{-5}$\,Mpc$^{-3}$, and in the HDF it is 
$2.5 \times 10^{-5}$\,Mpc$^{-3}$. 
}
\end{deluxetable}


\end{document}